

\def\singlespace{\normalbaselines}
\def\oneandahalfspace{\baselineskip=1.15\normalbaselineskip plus 1pt
\lineskip=2pt\lineskiplimit=1pt}

\def\np{\vfill\eject}
\def\nl{\hfil\break}

\def\nofirstpagenoten{\nopagenumbers\footline={\ifnum\pageno>1\tenrm
\hss\folio\hss\fi}}
\def\nofirstpagenotwelve{\nopagenumbers\footline={\ifnum\pageno>1\twelverm
\hss\folio\hss\fi}}
\def\leaderfill{\leaders\hbox to 1em{\hss.\hss}\hfill}

\def\frac#1/#2{\leavevmode\kern.1em
\raise.5ex\hbox{\the\scriptfont0 #1}\kern-.1em/\kern-.15em
\lower.25ex\hbox{\the\scriptfont0 #2}}
\def\sfrac#1/#2{\leavevmode\kern.1em
\raise.5ex\hbox{\the\scriptscriptfont0 #1}\kern-.1em/\kern-.15em
\lower.25ex\hbox{\the\scriptscriptfont0 #2}}


\parindent=20pt
\def\narrow{\advance\leftskip by 40pt \advance\rightskip by 40pt}

\def\AB{\bigskip
        \centerline{\bf \tenfoot ABSTRACT}\medskip\narrow}
\def\nonarrower{\advance\leftskip by -40pt\advance\rightskip by -40pt}
\def\AE{\bigskip\nonarrower}

\def\boxit#1{\vbox{\hrule\hbox{\vrule\kern3pt
        \vbox{\kern3pt#1\kern3pt}\kern3pt\vrule}\hrule}}

\def\gtorder{\mathrel{\raise.3ex\hbox{$>$}\mkern-14mu
             \lower0.6ex\hbox{$\sim$}}}
\def\ltorder{\mathrel{\raise.3ex\hbox{$<$}|mkern-14mu
             \lower0.6ex\hbox{\sim$}}}
\def\dalemb#1#2{{\vbox{\hrule height .#2pt
        \hbox{\vrule width.#2pt height#1pt \kern#1pt
                \vrule width.#2pt}
        \hrule height.#2pt}}}

\font\fourteentt=cmtt10 scaled \magstep2
\font\fourteenbf=cmbx12 scaled \magstep1
\font\fourteenrm=cmr12 scaled \magstep1
\font\fourteeni=cmmi12 scaled \magstep1
\font\fourteenss=cmss12 scaled \magstep1
\font\fourteensy=cmsy10 scaled \magstep2
\font\fourteensl=cmsl12 scaled \magstep1
\font\fourteenex=cmex10 scaled \magstep2
\font\fourteenit=cmti12 scaled \magstep1
\font\twelvett=cmtt10 scaled \magstep1 \font\twelvebf=cmbx12
\font\twelverm=cmr12 \font\twelvei=cmmi12
\font\twelvess=cmss12 \font\twelvesy=cmsy10 scaled \magstep1
\font\twelvesl=cmsl12 \font\twelveex=cmex10 scaled \magstep1
\font\twelveit=cmti12
\font\tenss=cmss10
 
 \font\ninebf=cmbx7 scaled \magstep1
\font\ninerm=cmr7 scaled \magstep1 \font\ninei=cmmi7 scaled \magstep1
\font\ninesy=cmsy7 scaled \magstep1 
\font\eightrm=cmr7 scaled 1140 
 
\font\sevenbf=cmbx7 \font\sevenrm=cmr7 \font\seveni=cmmi7
\font\sevensy=cmsy7 

\catcode`@=11
\newskip\ttglue
\newfam\ssfam

\def\fourteenpoint{\def\rm{\fam0\fourteenrm}
\textfont0=\fourteenrm \scriptfont0=\tenrm \scriptscriptfont0=\sevenrm
\textfont1=\fourteeni \scriptfont1=\teni \scriptscriptfont1=\seveni
\textfont2=\fourteensy \scriptfont2=\tensy \scriptscriptfont2=\sevensy
\textfont3=\fourteenex \scriptfont3=\fourteenex \scriptscriptfont3=\fourteenex
\def\it{\fam\itfam\fourteenit} \textfont\itfam=\fourteenit
\def\sl{\fam\slfam\fourteensl} \textfont\slfam=\fourteensl
\def\bf{\fam\bffam\fourteenbf} \textfont\bffam=\fourteenbf
\scriptfont\bffam=\tenbf \scriptscriptfont\bffam=\sevenbf
\def\tt{\fam\ttfam\fourteentt} \textfont\ttfam=\fourteentt
\def\ss{\fam\ssfam\fourteenss} \textfont\ssfam=\fourteenss
\tt \ttglue=.5em plus .25em minus .15em
\normalbaselineskip=16pt
\abovedisplayskip=16pt plus 4pt minus 12pt
\belowdisplayskip=16pt plus 4pt minus 12pt
\abovedisplayshortskip=0pt plus 4pt
\belowdisplayshortskip=9pt plus 4pt minus 6pt
\parskip=5pt plus 1.5pt
\setbox\strutbox=\hbox{\vrule height12pt depth5pt width0pt}
\let\sc=\tenrm
\let\big=\fourteenbig \normalbaselines\rm}
\def\fourteenbig#1{{\hbox{$\left#1\vbox to12pt{}\right.\n@space$}}}

\def\twelvepoint{\def\rm{\fam0\twelverm}
\textfont0=\twelverm \scriptfont0=\ninerm \scriptscriptfont0=\sevenrm
\textfont1=\twelvei \scriptfont1=\ninei \scriptscriptfont1=\seveni
\textfont2=\twelvesy \scriptfont2=\ninesy \scriptscriptfont2=\sevensy
\textfont3=\twelveex \scriptfont3=\twelveex \scriptscriptfont3=\twelveex
\def\it{\fam\itfam\twelveit} \textfont\itfam=\twelveit
\def\sl{\fam\slfam\twelvesl} \textfont\slfam=\twelvesl
\def\bf{\fam\bffam\twelvebf} \textfont\bffam=\twelvebf
\scriptfont\bffam=\ninebf \scriptscriptfont\bffam=\sevenbf
\def\tt{\fam\ttfam\twelvett} \textfont\ttfam=\twelvett
\def\ss{\fam\ssfam\twelvess} \textfont\ssfam=\twelvess
\tt \ttglue=.5em plus .25em minus .15em
\normalbaselineskip=14pt
\abovedisplayskip=14pt plus 3pt minus 10pt
\belowdisplayskip=14pt plus 3pt minus 10pt
\abovedisplayshortskip=0pt plus 3pt
\belowdisplayshortskip=8pt plus 3pt minus 5pt
\parskip=3pt plus 1.5pt
\setbox\strutbox=\hbox{\vrule height10pt depth4pt width0pt}
\let\sc=\ninerm
\let\big=\twelvebig \normalbaselines\rm}
\def\twelvebig#1{{\hbox{$\left#1\vbox to10pt{}\right.\n@space$}}}

\def\tenpoint{\def\rm{\fam0\tenrm}
\textfont0=\tenrm \scriptfont0=\sevenrm \scriptscriptfont0=\fiverm
\textfont1=\teni \scriptfont1=\seveni \scriptscriptfont1=\fivei
\textfont2=\tensy \scriptfont2=\sevensy \scriptscriptfont2=\fivesy
\textfont3=\tenex \scriptfont3=\tenex \scriptscriptfont3=\tenex
\def\it{\fam\itfam\tenit} \textfont\itfam=\tenit
\def\sl{\fam\slfam\tensl} \textfont\slfam=\tensl
\def\bf{\fam\bffam\tenbf} \textfont\bffam=\tenbf
\scriptfont\bffam=\sevenbf \scriptscriptfont\bffam=\fivebf
\def\tt{\fam\ttfam\tentt} \textfont\ttfam=\tentt
\def\ss{\fam\ssfam\tenss} \textfont\ssfam=\tenss
\tt \ttglue=.5em plus .25em minus .15em
\normalbaselineskip=12pt
\abovedisplayskip=12pt plus 3pt minus 9pt
\belowdisplayskip=12pt plus 3pt minus 9pt
\abovedisplayshortskip=0pt plus 3pt
\belowdisplayshortskip=7pt plus 3pt minus 4pt
\parskip=0.0pt plus 1.0pt
\setbox\strutbox=\hbox{\vrule height8.5pt depth3.5pt width0pt}
\let\sc=\eightrm
\let\big=\tenbig \normalbaselines\rm}
\def\tenbig#1{{\hbox{$\left#1\vbox to8.5pt{}\right.\n@space$}}}
\let\rawfootnote=\footnote \def\footnote#1#2{{\rm\parskip=0pt\rawfootnote{#1}
{#2\hfill\vrule height 0pt depth 6pt width 0pt}}}

\def\tenfoot{\tenpoint\hskip-\parindent\hskip-.1cm}

\overfullrule=0pt
\twelvepoint
\def\sbullet{\raise.2em\hbox{$\scriptscriptstyle\bullet$}}
\nofirstpagenotwelve
\hsize=16.5 truecm
\baselineskip 15pt

\def\noverm#1#2{{\textstyle{#1\over #2}}}
\def\half{\noverm{1}{2}}

\oneandahalfspace
\rightline{Preprint-KUL-TF-92/37}
\rightline{September 1992}

\vskip 2truecm
\centerline{\bf Generalised Virasoro Constructions}
\centerline{\bf from}
\centerline{\bf Affine In\"on\"u-Wigner Contractions}
\vskip 1.2truecm
\centerline{S. Schrans\footnote{$^\diamond$}{\tenfoot
Onderzoeker I.I.K.W.; e-mail: stschr@iks.kuleuven.ac.be.\nl
Address after March 1, 1993:
Koninklijke/Shell-Laboratorium Amsterdam (Shell Research B.V.),\nl
Badhuisweg 3, 1031 CM Amsterdam, The
Netherlands.}
}
\vskip 1.2truecm
\centerline{\it Instituut voor Theoretische Fysica}
\centerline{\it K.U.Leuven}
\centerline{\it Celestijnenlaan 200D}
\centerline{\it B-3001 Leuven}
\centerline{\it Belgium}

\vskip 1.2truecm
\AB\singlespace
We present a new method to find solutions of the Virasoro master equations for
any  affine Lie algebra $\widehat{g}$. The basic idea is to consider first the
simplified case of an In\"on\"u-Wigner contraction $\widehat{g}_c$ of
$\widehat{g}$ and to extend the Virasoro constructions of $\widehat{g}_c$ to
$\widehat{g}$ by a perturbative expansion in the contraction parameter.  The
method is then applied to the orthogonal algebras, leading to  fixed-level
multi-parameter Virasoro constructions, which are the generalisations of the
one-parameter Virasoro construction of $\widehat{su}(2)$ at level four.
\AE\oneandahalfspace

\vskip 1.2truecm
\centerline{\tenfoot Available from hepth@xxx/9209090}


\def\reg{{\rm regular}}

\np
\noindent {\bf 1.~Introduction}
\bigskip
\noindent An important ingredient in the study of two-dimensional conformal
field theories is the realisation of their chiral symmetry algebras in terms of
simpler structures such as free fields or affine currents. The interest in
realisations of the two-dimensional conformal symmetry algebra, the Virasoro
algebra, on affine Lie algebras has increased considerably since it has been
noted that the Sugawara and coset constructions are but specific examples of
far more general ``Virasoro constructions.'' The most general realisations of
the Virasoro algebra as a quadratic combination
$$
T(z) = L_{ab} (J^aJ^b)(z), \eqno(1)$$
with $ L_{ab}=L_{ba}$, of affine currents $J^a(z)$, generating the affine Lie
algebra $\widehat{g}$
$$
J^a(z)J^b(w) = {\ell g^{ab}\over (z-w)^2} + {i {f^{ab}}_c
J^c(w)\over z-w} + \reg ,
\eqno(2)$$
leads to a set of coupled algebraic equations [1,2], the so-called Virasoro
master  equations for $\widehat{g}$,
$$
L_{ab} = 2 \ell L_{ac}g^{cd} L_{db} - {f^{kl}}_a {f^{cd}}_b L_{kc}L_{ld} -
{f^{ld}}_c{f^{kc}}_{(a}L_{b)d} L_{kl},
\eqno(3)$$
where the brackets denote symmetrisation, $X_{(ab)}=X_{ab}+X_{ba}$. The central
charge of the resulting Virasoro algebra is then given by
$$
c= 2 \ell g^{ab}L_{ab}.
\eqno(4)$$
An important property of the Virasoro master equations for (products of) simple
algebras is that all its solutions come in ``conjugate pairs:'' if $L_{ab}$ is
a solution (with central charge $c$), then so  is $L^{\rm Sug}_{ab}- L_{ab}$
(with central charge $c_{\rm Sug}-c$), where  $L^{\rm Sug}_{ab}$ denotes the
Sugawara solution (and $c_{\rm Sug}$ the Sugawara central charge).

Even though only quadratic the Virasoro master equations (3) have so far
resisted any attempt at a general solution. In fact the only simple algebra for
which the complete solution is known is $\widehat{su}(2)$ [2]. For all other
algebras one has to recur to specific ansatze in order to reduce the number of
nonlinear equations in such a way that the resulting simplified set can be
solved. Such approaches have led to numerous new Virasoro constructions (see
{\it e.g.} [1--7]). So far, the only method leading to a systematic treatment
of these ``simplified master equations'' is a perturbative expansion,
developped by Halpern and Obers, in the inverse level [4]. Solutions of the
zeroth order equations, which  reduce to constructions on (products of)
$\widehat{u}(1)$ algebras, can be straightforwardly extended to higher order,
often leading to  new constructions. This ``high-level expansion'' has, in
particular, been very succesful in the case of  orthogonal algebras, leading to
a surprising connection with graph theory [5]. The major drawback of this
perturbative method, however, is that, by construction, solutions which exist
only for fixed values of the level of the affine algebra cannot be obtained.
The canonical example of such a fixed-level Virasoro construction is the only
$\widehat{su}(2)$ solution that is not a Sugawara or coset construction. It
occurs at level four, contains a free parameter and has central charge $c=1$.

In this letter  we present a new perturbative approach which gives solutions
for generic as well as for specific values of the level. The starting point is
to solve the master equations for an In\"on\"u-Wigner contraction of the
original algebra and then extend these solutions to the original algebra by
expanding in the contraction parameter.

In the next section, we shall illustrate the basic ideas on the
$\widehat{su}(2)$ example, recovering the one-parameter solution at level four.
In the third section we discuss the general case and derive a necessary
condition for the existence of fixed-level solutions. In the fourth section we
show that this necessary condition is also sufficient for orthogonal algebras
and find fixed-level multi-parameter Virasoro constructions.  More
specifically, we find for any $\widehat{so}(n)$ algebra $n-2$ solutions with
integer central charges $c=k$, $k=1,2,\ldots,n-2$, containing $(k-1)(n-k)$
arbitrary (complex) parameters. Finally, in the last section we present some
conclusions.

\bigskip\bigskip
\noindent {\bf 2. An illustrative example}
\bigskip
\noindent Consider as an example the case $\widehat{g}=\widehat{su}(2) \cong
\widehat{so}(3)$ generated by $J^a(z), a=1,2,3$. Take $g^{ab}=\half
\delta^{ab}$ and ${f^{ab}}_c= \varepsilon_{abc}$ so that $\ell$ is the level of
the affine algebra. In this case, the most general Virasoro constructions  are
given by the diagonal ones: $L_{ab}= \lambda_a \delta_{ab}$ [1,2]. Even though
it is a simple  exercise to solve the corresponding master equations
completely, we shall only do so indirectly.

Perform the transformation
$$\eqalign{
J^{1,2}(z) &\rightarrow \sqrt{\epsilon}\, J^{1,2}(z),\cr
J^3(z) &\rightarrow \;\,J^3(z).
\cr}\eqno(5)$$
The limit $\epsilon \rightarrow 0$ is an In\"on\"u-Wigner contraction [8] and
leads to the affine two-dimensional Euclidean algebra $\widehat{so}(2)
\otimes \widehat{T}_2 \equiv \widehat{e}(2)$.
The master equations are  after this rescaling
$$\eqalign{
\lambda_1 &= \epsilon\ell \lambda_1^2 + 2 \epsilon\lambda_1\lambda_2 +
2\lambda_1 \lambda_3 - 2 \lambda_2 \lambda_3,\cr
\lambda_2 &= \epsilon\ell \lambda_2^2 + 2 \epsilon\lambda_1\lambda_2 +
2\lambda_2 \lambda_3 - 2 \lambda_1 \lambda_3,\cr
\lambda_3 &= \ell \lambda_3^2 + 2 \epsilon\lambda_1\lambda_3 +
2\epsilon\lambda_2 \lambda_3 - 2\epsilon^2 \lambda_1 \lambda_2.
\cr}\eqno(6)$$
Expanding
$$
\lambda_a = \sum_{s=0}^{\infty} \epsilon^s\, \lambda_a^{(s)},
\eqno(7)$$
leads, at order zero, to the master equations for $\widehat{e}(2)$:
$$\eqalignno{
\lambda_1^{(0)} &= 2
\lambda_3^{(0)}(\lambda_1^{(0)}-\lambda_2^{(0)}),&(8)\cr
\lambda_2^{(0)} &= 2
\lambda_3^{(0)}(\lambda_2^{(0)}-\lambda_1^{(0)}),&(9)\cr
\lambda_3^{(0)} &= \ell \left(\lambda_3^{(0)}\right)^2 .
&(10)\cr}$$
These equations are clearly easier to solve than the corresponding ones for
$\widehat{su}(2)$. The solution $\lambda_3^{(0)}=0$ from (10) leads to the
trivial solution $\lambda_1^{(0)}=\lambda_2^{(0)}=0$. The other solution of
(10), $\lambda_3^{(0)}=1/\ell$, implies in general
$\lambda_1^{(0)}=\lambda_2^{(0)}=0$, except when $\ell=4$, when
$\lambda_1^{(0)}=-\lambda_2^{(0)}=\alpha_0$ is a solution for an arbitrary
parameter $\alpha_0$.

The higher-order equations in $\epsilon$ are now all linear and hence
straightforward to solve. The trivial solution at zeroth order remains trivial
at all orders, while the $\lambda_1^{(0)}=\lambda_2^{(0)}=0$, $
\lambda_3^{(0)}=1/\ell$ solution for generic $\ell$ gives
$\lambda_i^{(n\geq1)}=0$ and thus  the  construction corresponding to the
energy-momentum tensor of the $\widehat{u}(1)$ subalgebra generated by
$J^3(z)$. Finally, the solution at level four has an  expansion
$$\eqalign{
\lambda_1 &=  \alpha_0 + \alpha_1\,\epsilon +  \alpha_2\,\epsilon^2+
\alpha_3\,\epsilon^3 +  \alpha_4\,\epsilon^4 + o(\epsilon^5),\cr
\lambda_2 &= -\alpha_0 +  (4 \alpha_0^2- \alpha_1)\,\epsilon +
(-16 \alpha_0^3 + 8 \alpha_0 \alpha_1 - \alpha_2)\,\epsilon^2\cr
&+
(128 \alpha_0^4 -48 \alpha_0^2 \alpha_1 + 4 \alpha_1^2 + 8 \alpha_0\alpha_2 -
\alpha_3)\,\epsilon^3\cr
& +   (-1024 \alpha_0^5 + 512 \alpha_0^3 \alpha_1 - 48 \alpha_0
\alpha_1^2 - 48 \alpha_0^2 \alpha_2 + 8 \alpha_1 \alpha_2 + 8 \alpha_0\alpha_3
-\alpha_4)\,\epsilon^4 + o(\epsilon^5),\cr
\lambda_3 &= {1\over 4} -4 \alpha_0^2\,\epsilon^2 +
(16 \alpha_0^3 -8 \alpha_0 \alpha_1)\,\epsilon^3\cr
& +  (-128 \alpha_0^4 + 48 \alpha_0^2 \alpha_1 - 4
\alpha_1^2 - 8 \alpha_0 \alpha_2)\,\epsilon^4 + o(\epsilon^5).
\cr}\eqno(11)$$
By redefining
$$
\alpha_0 = \tilde\alpha_0 - \alpha_1\,\epsilon -  \alpha_2\, \epsilon^2
-  \alpha_3\,\epsilon^3-  \alpha_4\,\epsilon^4 - o(\epsilon^5)
\eqno(12)$$
this can be rewritten as
$$\eqalign{
\lambda_1 &=  \tilde\alpha_0,\cr
\lambda_2 &= -\tilde\alpha_0 +  4 \tilde\alpha_0^2  \,\epsilon
-16 \tilde\alpha_0^3\,\epsilon^2 +128 \tilde\alpha_0^4\,\epsilon^3
-1024 \tilde\alpha_0^5\,\epsilon^4+ o(\epsilon^5),\cr
\lambda_3 &= {1\over 4}
-4 \tilde\alpha_0^2 \,\epsilon^2+16 \tilde\alpha_0^3\,\epsilon^3
-128 \tilde\alpha_0^4\,\epsilon^4 + o(\epsilon^5),
\cr}\eqno(13)$$
which explicitly shows that there is only one relevant parameter.

This solution satisfies
$$\eqalign{
\epsilon ( \lambda_1 + \lambda_2) + \lambda_3 &= {1\over 4} + o(\epsilon^5),
\cr
\epsilon^2 ( \lambda_1^2 + \lambda_2^2) + \lambda_3^2 &= {1\over 16} +
o(\epsilon^5)\cr}
\eqno(14)$$
and has central charge $c=1 + o(\epsilon^5)$. It is precisely the one-parameter
$\widehat{su}(2)$ Virasoro construction at level four.

\bigskip\bigskip
\noindent {\bf 3. The general case}
\bigskip
\noindent Consider an affine Lie algebra $\widehat{g}$ with generators
$J^a(z)$. Split the generators of $\widehat{g}$ into two sets  as follows
$$
\{J^a(z)\} =  \{J^\alpha(z)\} \cup \{J^A(z)\}
\eqno(15)$$
and perform the scale transformation
$$\eqalign{
J^\alpha(z) &\rightarrow \;\,J^\alpha(z),\cr
J^A(z) &\rightarrow \epsilon\, J^A(z).
\cr}\eqno(16)$$
The structure constants then rescale as
$$\eqalign{
{f^{AB}}_C\rightarrow \epsilon\, {f^{AB}}_C, &\quad \quad \quad  \quad
{f^{AB}}_\gamma\rightarrow \epsilon^2\,{f^{AB}}_\gamma, \cr
{f^{A\beta}}_C\rightarrow {f^{A\beta}}_C, &\quad \quad \quad \quad
{f^{A\beta}}_\gamma\rightarrow \epsilon \,{f^{A\beta}}_\gamma, \cr
{f^{\alpha\beta}}_C\rightarrow \epsilon^{-1}\,{f^{\alpha\beta}}_C, &
\quad \quad \quad \quad
{f^{\alpha\beta}}_\gamma\rightarrow {f^{\alpha\beta}}_\gamma
\cr}\eqno(17)$$
and the Killing form as
$$
g^{AB}\rightarrow \epsilon^2\, g^{AB}, \quad\quad
g^{A\alpha}\rightarrow \epsilon\, g^{A\alpha}, \quad\quad
g^{\alpha\beta} \rightarrow g^{\alpha\beta}.
\eqno(18)$$
Existence of the limit $\epsilon\rightarrow 0$  requires
${f^{\alpha\beta}}_C=0$, or thus the currents $J^\alpha(z)$ to form a
subalgebra $\widehat{h}$. This limit then corresponds to an In\"on\"u-Wigner
contraction [8] $\widehat{g}_c$ of $\widehat{g}$.\footnote{$^*$}{\tenfoot This
is equivalent to considering the affinisation of the contraction of the finite
dimensional algebra $g$, see [9].}

We shall expand $L_{ab}$ and $c$ in powers of $\epsilon$
$$
L_{ab}= \sum_{s=0}^\infty \epsilon^s\, L_{ab}^{(s)}, \quad\quad
c= \sum_{s=0}^\infty \epsilon^s\, c^{(s)}.
\eqno(19)$$
The zeroth order equations then reduce to the master equations for the
contracted algebra $\widehat{g}_c$. Even though still quadratic, they are
easier to solve than the corresponding equations for $\widehat{g}$, since at
that order one can put ${f^{AB}}_C={f^{AB}}_\gamma={f^{A \beta}}_\gamma=0$ and
$g^{AB}=g^{A\beta}=0$. The higher order equations now all become linear and are
hence straightforward to solve order by order.

Instead of writing the general form of these equations down, we shall only
consider the ``natural'' ansatz
$$
L_{\alpha B}^{(0)} = 0,
\eqno(20)$$
which is sufficient for our purposes. The zeroth order master equations then
become
$$\eqalignno{
L^{(0)}_{\mu\nu} &= 2 \ell L^{(0)}_{\mu\gamma}g^{\gamma\delta}
L^{(0)}_{\delta\nu} -
{f^{\alpha\beta}}_\mu {f^{\gamma\delta}}_\nu L^{(0)}_{\alpha\gamma}
L^{(0)}_{\beta\delta} -
{f^{\beta\delta}}_\gamma {f^{\alpha\gamma}}_{(\mu}L^{(0)}_{\nu)\delta}
L^{(0)}_{\alpha\beta},&(21)\cr
L^{(0)}_{MN}&= L^{(0)}_{\alpha\gamma}{f^{B\alpha}}_{(M} \left( L^{(0)}_{N)D}
{f^{\gamma D}}_B +
{f^{\gamma D}}_{N)} L^{(0)}_{DB} \right).
&(22)\cr}$$
The equations (21) are precisely the master equations for the subalgebra
$\widehat{h}$. Once a solution for $\widehat{h}$ has been chosen, (22) become
homogeneous linear equations in the remaining variables $L^{(0)}_{MN}$. A
nontrivial solution exists only if the determinant of the coefficient matrix is
zero. Requiring this determinant to vanish leads to a polynomial equation in
the level, whose roots provide candidate values of the level for which
continuous Virasoro constructions exist for $\widehat{g}$.

Let us work this out in more detail for a specific example.  Suppose that $g$
is simple and choose a specific solution $L_{\alpha\beta}^{(0)}$ for (21).
Assume the diagonal ansatz
$$
L_{AB}= g_{AB} \lambda_B \quad\quad\quad {\rm (no\  sum)}.
\eqno(23)$$
After multiplication by $g^{NM}$ and summation  over $N$, (22) can be
rewritten as ${\cal M}_M^P \lambda_P =0$ where we have defined the matrix
${\cal M}$ as
$$
{\cal M}_M^P=\sum_N g_{MN}g^{NM}\delta_M^P +
2 \sum_{N,B,\alpha,\gamma} L^{(0)}_{\alpha\gamma} g^{NM}{f^{\alpha B}}_M
\left({f^{\gamma P}}_B
g_{NP}+ {f^{\gamma P}}_N g_{PB}\right).
\eqno(24)$$
(For convenience, we have written out all the summation indices explicitly.)
A necessary condition for a solution is hence
$$
{\rm det}({\cal M})=0.
\eqno(25)$$
Note that for symmetric spaces $\sum_N g_{AN} g^{NB} = \delta_A^B$. In the next
section we shall consider specific ansatze for orthogonal algebras  where  this
condition will turn out to be sufficient as well.

\bigskip\bigskip
\noindent {\bf 4. Orthogonal algebras}
\bigskip
\noindent Consider the orthogonal affine Lie algebra $\widehat{g}=
\widehat{so}(n)$. Choose as a basis  the currents $J^{a\bar{a}}(z)$ with vector
indices
$1\leq a < \bar{a} \leq n$ and operator product expansions
$$
J^{a\bar{a}}(z)J^{b\bar{b}}(w)= {\ell g^{a\bar{a},b\bar{b}} \over (z-w)^2} +
{i {f^{a\bar{a},b\bar{b}}}_{c\bar{c}} J^{c\bar{c}} \over z-w} + \reg,
\eqno(26)$$
where the structure constants are
$$
{f^{a\bar{a},b\bar{b}}}_{c\bar{c}}=
\delta^{\bar{a}b}\delta^{a}_{[c}\delta^{\bar{b}}_{\bar{c}]} +
\delta^{a\bar{b}}\delta^{\bar{a}}_{[c}\delta^{b}_{\bar{c}]} -
\delta^{ab}\delta^{\bar{a}}_{[c}\delta^{\bar{b}}_{\bar{c}]} -
\delta^{\bar{a}\bar{b}}\delta^{a}_{[c}\delta^{b}_{\bar{c}]}
\eqno(27)$$
and the Killing form is
$$
g^{a\bar{a},b\bar{b}}= \kappa_n \delta^{a[b} \delta^{\bar{b}]\bar{a}},
\eqno(28)$$
the square brackets denoting antisymmetrisation, $X_{[ab]}=X_{ab}-X_{ba}$.
Taking the normalisation $\kappa_3=1/2$ and $ \kappa_{n\geq 4}=1$,
fixes $\ell$ as the level of the $\widehat{so}(n)$ algebra.
The quadratic combination
$$
T(z)= L_{a\bar{a},b\bar{b}} (J^{a\bar{a}}J^{b\bar{b}})(z), \quad\quad\quad
L_{a\bar{a},b\bar{b}}=L_{b\bar{b},a\bar{a}},
\eqno(29)$$
with the diagonal ansatz
$$
L_{a\bar{a},b\bar{b}}= L_{a\bar{a}} \delta_{a[b}\delta_{\bar{b}]\bar{a}},
\eqno(30)$$
generates a Virasoro algebra provided [5]
$$
L_{ab}= 2 \kappa_n \ell L_{ab}^2 + 2 \sum_{c\neq a,b}^n \left[
L_{ab}(L_{ac}+ L_{bc})- L_{ac}L_{bc}\right].
\eqno(31)$$
The central charge is then given by
$$
c= 2 \ell \kappa_n \sum_{a<b}^n L_{ab}.
\eqno(32)$$
Consider now a subalgebra $\widehat{h}=\widehat{so}(k)$ of $\widehat{so}(n)$
generated by the currents $J^{\alpha\bar{\alpha}}(z)$ with $1\leq \alpha <
\bar{\alpha}\leq k<n$. Rescale the currents as
$$\eqalign{
J^{\alpha\bar{\alpha}}(z)&\rightarrow J^{\alpha\bar{\alpha}}(z), \cr
J^{\alpha B}(z)&\rightarrow \sqrt{\epsilon}\, J^{\alpha B}(z), \cr
J^{C\bar{C}}(z) &\rightarrow \sqrt{\epsilon}\, J^{C\bar{C}}(z),
\cr}\eqno(33)$$
with $k+1\leq B \leq n$ and $k+1\leq C <\bar{C}\leq n$. Since the currents
$J^{\alpha\bar{\alpha}}(z)$ generate a subalgebra, the limit
$\epsilon\rightarrow 0$ exists. When $k=n-1$ this  yields  $\widehat{e}(n-1)$,
the affine $(n-1)$-dimensional Euclidean algebra.  As usual we expand
$$
L_{ab} = \sum_{s=0}^\infty \epsilon^s\, L_{ab}^{(s)} \, .
\eqno(34)$$

As explained in the previous section, we can choose  for
$L_{\alpha\beta}^{(0)}$ an arbitrary solution of the Virasoro master equations
for $\widehat{so}(k)$. In this letter we shall consider the Sugawara
construction:
$$
L_{\alpha\beta}^{(0)}= {1 \over 2 \kappa_n(\ell+k-2)},\quad\quad
{\rm for\ all\ } 1\leq \alpha <\beta \leq k.
\eqno(35)$$
Plugging (27), (28), (30) and (35) into (24) gives the following simple result
for the matrix ${\cal M}$:
$$
{\cal M}_{\alpha A}^{\beta B}= \left\{ \left[ 1- {k\over\kappa_n
(\ell+ k-2)}\right]
\delta_\alpha^\beta + {1\over \kappa_n (\ell+ k-2)}\right\}\delta_A^B.
\eqno(36)$$
The determinant of this matrix can be easily computed:
$$
{\rm det}({\cal M}) = \left[1 - {k\over \kappa_n (\ell + k-2)}
\right]^{(k-1)(n-k)}.
\eqno(37)$$
Hence we deduce that possible continuous Virasoro constructions might arise
at fixed level
$$
\ell= 2 + {1-\kappa_n\over\kappa_n} k,
\eqno(38)$$
or thus for $\widehat{so}(3)$  at level 4 and for $\widehat{so}(n\geq 4)$ at
level 2. More precisely, (37) suggests that, given $n$, there might be $n-2$
different constructions corresponding to a specific choice of a subalgebra
$\widehat{so}(k)$, $k=2,3,\ldots,n-1$, each one having  $(k-1)(n-k)$ arbitrary
parameters.

A detailed analysis of the higher-order equations indicates that all the  above
constructions do in fact exist! Even though we have no proof of this statement
in the general case, we have performed extensive checks (up to sixth order in
the expansion parameter and up to and including $\widehat{so}(10)$) and are
confident that {\it all} these constructions are valid to all orders in
$\epsilon$. Since these perturbative expansions are quite complicated and are
not very enlightening by themselves, we shall refrain from giving them
explicitly and shall only comment on some of their main characteristics.

For $\widehat{so}(3)_4$ we  recover exactly  the construction from section 2
after the identification $J^{23}(z)\equiv J^1(z)$, $J^{13}(z)\equiv J^2(z)$ and
$J^{12}(z)\equiv J^3(z)$. The two construction for $\widehat{so}(4)$ have
already been conjectured to exist [6]. They both contain two arbitrary
parameters as expected. Using the form of the perturbative expansion, we have
been able to find the explicit nonperturbative form of both these
constructions. This result is presented in the appendix. For
$\widehat{so}(n\geq 5)$ we do not have a closed-form expression for the
constructions, but, as already mentioned, sufficient evidence from the
perturbative expansion. Again for all these constructions the number of
arbitrary parameters is precisely the one indicated in (37). All these
parameters are nontrivial.

Thus the vanishing of the determinant (37) is a sufficient condition for the
existence of fixed-level multi-parameter Virasoro constructions. Let us, for
definiteness, denote by ${\cal V}(n,k)$ ($k=2,3,\ldots, n-1$) the construction
for $\widehat{g}= \widehat{so}(n)$ which arises from the
$\widehat{h}=\widehat{so}(k)$  subalgebra. ${\cal V}(n,k)$ thus contains
$(k-1)(n-k)$ arbitrary (complex) parameters. Moreover (up to the order of the
perturbative expansion that was checked and exactly for the $n=3$ and $n=4$
cases) the central charge of  ${\cal V}(n,k)$ is $c_{{\cal V}(n,k)}=k-1$. At
the precise level at which these Virasoro constructions occur, the Sugawara
construction for $\widehat{so}(n)$ has integer central charge $c_n= n-1$.
Since
$$
c_{{\cal V}(n,k)}+c_{{\cal V}(n,n+1-k)}= n-1 = c_n
\eqno(39)$$
and moreover ${\cal V}(n,k)$ and ${\cal V}(n,n+1-k)$ contain the same number of
parameters, it is natural to conjecture that they form conjugate pairs. This
implies in particular that ${\cal V}(2p+1,p+1)$ is closed under
conjugation.\footnote{$^*$}{\tenfoot  Two-parameter constructions for
$\widehat{so}(2p+1)$, $p\geq 2$ with central charge $c=p$ have already been
described in [5] (where they were called $SO(2p+1)_2^\sharp [d,6]$); these
constructions are closed under conjugation. It is conceivable that they form a
special case of ${\cal V}(2p+1,p+1)$, which contains $p^2$ parameters.}

\bigskip\bigskip
\noindent {\bf 5. Conclusions}
\bigskip
\noindent In this letter we have presented a new perturbative method, based on
In\"on\"u-Wigner contractions of affine Lie algebras, to find solutions of
the Virasoro master equations. The major advantage of our method is that,
contrary to the high-level analysis of Halpern and Obers [4], our method seems
ideally suited to investigate the existence of fixed-level continuous Virasoro
constructions. We have illustrated this on orthogonal algebras, resulting in a
class of fixed-level multi-parameter deformations of coset and Sugawara
constructions. These constructions are the natural generalisations of the
one-parameter $\widehat{su}(2)_4$ construction.

While completing this letter we encountered a preprint, [10], where these
fixed-level multi-parameter orthogonal constructions are derived as
deformations of coset constructions for $\widehat{so}(n)/\widehat{so}(k)$. The
agreement on overlapping results is perfect.

Even though we have focussed in this letter on fixed-level constructions it is
clear that our method applies as well to Virasoro constructions existing for
generic values of the level. The major drawback of our method is, however, that
it does not lead to  an exhaustive list of these solutions (see section 2), so
that it might be better suited for the search of new fixed-level constructions.
This search, together with other applications, based on {\it e.g.} double
In\"on\"u-Wigner contractions, is under investigation.

\bigskip\bigskip
\noindent {\bf Appendix}
\bigskip
\noindent Here we present as an example the nonperturbative form of the two
solutions for $\widehat{g}=\widehat{so}(4)_2$.
The first one arises from
$\widehat{h}=\widehat{so}(2)$ and contains 2 free (complex) parameters $\alpha$
and $\beta$. It
is given by
$$
L_{12}= x_1, \quad\quad\quad L_{13}=x_2, \quad\quad\quad L_{23}=x_3,
$$
$$
L_{14}={1\over 4} - x_1-x_2-x_3-{\alpha}-{\beta}, \quad\quad
L_{24}={\beta}, \quad\quad L_{34}={\alpha}.
\eqno(40)$$
Here
$$
x_3={- {\alpha}{\beta} [1+4({\alpha}+{\beta}) -
\Delta]\over 8({\alpha}^2+{\alpha}{\beta}+{\beta}^2)},
\eqno(41)$$
with $\Delta$ given by
$$
\Delta= \pm \sqrt{1+ 8(\alpha+\beta) -48(\alpha^2+\beta^2)-32\alpha\beta}
\eqno(42)$$
and $x_1$ and $x_2$ by
$$
x_1={{A}({\alpha},{\beta})+ {B}({\alpha},
{\beta})\Delta \over {\cal N}({\alpha},{\beta})},
\quad\quad\quad
x_2={{A}({\beta},{\alpha})+ {B}({\beta},
{\alpha})\Delta \over {\cal N}({\beta},{\alpha})},
\eqno(43)$$
with
$$\eqalign{
{\cal N}({\alpha},{\beta}) &= 8({\alpha}^2+
{\alpha}{\beta}+{\beta}^2)[-2 \beta(\alpha+\beta) +
4\beta(\alpha^2+\beta^2)-\alpha^2 + {\beta}({\alpha}+
{\beta}) \Delta],\cr
{A}({\alpha},{\beta}) &= {\beta} \{(4{\alpha}-1)
[{\alpha}^3+2{\alpha}(16{\alpha}+3)
{\beta}^2] + 4 {\alpha}^2(-1+{\alpha}+8{\alpha}^2)
{\beta} \cr
&+(-3-12{\alpha}+128{\alpha}^2){\beta}^3 +
4 (1+32{\alpha}){\beta}^4 + 32 {\beta}^5\},\cr
{B}({\alpha},{\beta})&={\beta}({\alpha}+
{\beta})[{\alpha}^2+ 3 {\beta}({\alpha}+
{\beta}) -8 {\beta} ({\alpha}^2+{\beta}^2)].
\cr}\eqno(44)$$
This Virasoro construction has central charge $c=1$. For some special values of
the parameters these construction reduce to Sugawara or coset constructions:
for $(\alpha,\beta)=(0,1/4)$ we recover the Sugawara construction for the
$\widehat{so}(2)$ subalgebra generated by $J^{24}(z)$,
$(\alpha,\beta)=(0,-1/12)$ gives a $\widehat{so}(3)/\widehat{so}(2)$ coset
construction, $(\alpha,\beta)=(1/8,-1/8)$ an $\widehat{so}(4)/ (\widehat{so}(2)
\times \widehat{so}(2))$ coset construction, {\it etc}.

The second construction for $\widehat{so}(4)_2$,  arising from
$\widehat{h}=\widehat{so}(3)$,  also contains two free
(complex) parameters and has central charge $c=2$. Explicitly
$$
L_{12}= \tilde{x}_1, \quad\quad\quad L_{13}=\tilde{x}_2,
\quad\quad\quad L_{23}=\tilde{x}_3,
$$
$$
L_{14}={1\over 2} - \tilde{x}_1-\tilde{x}_2-\tilde{x}_3-
\tilde{\alpha}-\tilde{\beta}, \quad\quad
L_{24}=\tilde{\beta}, \quad\quad L_{34}=\tilde{\alpha}.
\eqno(45)$$
Here $\tilde{\alpha
}$ and $\tilde{\beta}$ are the free (complex) parameters and
$$
\tilde{x}_3={(8\tilde{\alpha}-1)(8\tilde{\beta}-1) \Delta - [4(\tilde{\alpha}+
\tilde{\beta})-1] [ 5 -24(\tilde{\alpha}+\tilde{\beta}) +
64 \tilde{\alpha}\tilde{\beta}] \over 8[3-24(\tilde{\alpha}+\tilde{\beta})
+ 64(\tilde{\alpha}^2+\tilde{\alpha}\tilde{\beta} + \tilde{\beta}^2)]},
\eqno(46)$$
with $\Delta$ given by (42),
and $\tilde{x}_1$ and $\tilde{x}_2$ are
$$
\tilde{x}_1={\widetilde{A}(\tilde{\alpha},\tilde{\beta})+
\widetilde{B}(\tilde{\alpha}, \tilde{\beta})\Delta \over \widetilde{\cal
N}(\tilde{\alpha},\tilde{\beta})}, \quad\quad\quad
\tilde{x}_2={\widetilde{A}(\tilde{\beta},\tilde{\alpha})+
\widetilde{B}(\tilde{\beta}, \tilde{\alpha})\Delta \over \widetilde{\cal
N}(\tilde{\beta},\tilde{\alpha})},
\eqno(47)$$
with
$$\eqalign{
\widetilde{\cal N}(\tilde{\alpha},\tilde{\beta}) &= 4[3-24(\tilde{\alpha}+
\tilde{\beta}) + 64(\tilde{\alpha}^2+\tilde{\alpha}
\tilde{\beta}+\tilde{\beta}^2)]
\{(8\tilde{\beta}-1)[4(\tilde{\alpha}+\tilde{\beta})-1]  \Delta \cr
&+2[1-6\tilde{\alpha} -8\tilde{\beta} + 8(1+8\tilde{\beta})(\tilde{\alpha}^2
+\tilde{\beta}^2) + 16\tilde{\alpha}\tilde{\beta}] \},\cr
\widetilde{A}(\tilde{\alpha},\tilde{\beta}) &= (1-4\tilde{\alpha})[1-32
\tilde{\alpha}+224\tilde{\alpha}^2-384\tilde{\alpha}^3 - 16(1-
28\tilde{\alpha}+160\tilde{\alpha}^2-128 \tilde{\alpha}^3)\tilde{\beta}\cr
&-1024(3-16\tilde{\alpha}+32\tilde{\alpha}^2) \tilde{\beta}^3]+ 16(17-288
\tilde{\alpha}+1792 \tilde{\alpha}^2-4096 \tilde{\alpha}^3+2048
\tilde{\alpha}^4)\tilde{\beta}^2\cr
&+ 512(33-192\tilde{\alpha}+256\tilde{\alpha}^2)\tilde{\beta}^4 + 8192(
16\tilde{\alpha}-5)\tilde{\beta}^5 + 32768\tilde{\beta}^6,\cr
\widetilde{B}(\tilde{\alpha},\tilde{\beta})&= -(8\tilde{\beta}-1)^2 [4(
\tilde{\alpha}+\tilde{\beta})-1][32(\tilde{\alpha}^2+\tilde{\beta}^2)-1].
\cr}\eqno(48)$$
Special points of interest are $(\alpha,\beta)=(1/8,-1/8)$ giving an
$\widehat{so}(4)/\widehat{so}(2)$ coset construction, $(\alpha,\beta)=
(1/8,5/24)$ an $\widehat{so}(4)/(\widehat{so}(3)/\widehat{so}(2))$ nested coset
construction, $(\alpha,\beta)=(0,1/4)$ an $\widehat{so}(2)\times
\widehat{so}(2)$ Sugawara construction, {\it etc}.

Conjugation with respect to $\widehat{so}(4)$ transforms these solutions into
one another.

\bigskip\bigskip
\centerline{\bf ACKNOWLEDGMENTS}
\bigskip

I am  grateful to Kris Thielemans for the unrelenting sharing  of  his
Mathematica expertise, to Marty Halpern for valuable e-discussions on the
orthogonal constructions and to Alex Deckmyn for his diligent reading of the
manuscript.

\bigskip\bigskip
\singlespace
\centerline{\bf REFERENCES}
\frenchspacing
\bigskip

\item{[1]}M.~B.~Halpern and E.Kiritsis, Mod.~Phys.~Lett.~{\bf A4} (1989) 1373;
erratum {\it ibid} {\bf A4} (1989) 1797.

\item{[2]}A.~Y.~Morozov, A.~M.~Perelomov, A.~A.~Rosly, M.~A.~Shifman and
A.~V.~Turbiner, Int.~Journ.~Mod.~Phys.~{\bf A5} (1990) 803.

\item{[3]}S.~Schrans and W.~Troost, Nucl.~Phys.~{\bf B345} (1990) 584.

\item{[4]}M.~B.~Halpern and N.~A.~Obers, Nucl.~Phys.~{\bf B345} (1990) 607.

\item{[5]}M.~B.~Halpern and N.~A.~Obers, Commun.~Math.~Phys.~{\bf 138}
(1991) 63.

\item{[6]}A.~Giveon, M.~B.~Halpern, E.~B.~Kiritsis and N.~A.~Obers,
Nucl.~Phys.~{\bf B357} (1991) 655.

\item{[7]}M.~B.~Halpern, in ``Strings and Symmetries 1991,'' ed.~N.~Berkovits
{\it et al}, World Scientific 1992.

\item{[8]}E.~In\"on\"u and E.~P.~Wigner, Proc.~Nat.~Acad.~Sci.~(US) {\bf 39}
(1953) 510.

\item{[9]}P.~Majumdar, {\sl In\"on\"u-Wigner Contraction of Kac-Moody
Algebras}, IMSc/92-26 (June 1992).

\item{[10]}A.~A.~Belov and Yu.~E.~Lozovik, Journ.~Nucl.~Phys.~(Yadernaya
fizika)  {\bf 53} (1991) 1464 (in Russian).

\bye